\documentclass[twocolumn,pra,preprintnumbers,amsmath,amssymb,superscriptaddress]{revtex4}
\usepackage{graphicx}
\pdfoutput=1

\newcommand{\bea}{\begin{eqnarray}}
\newcommand{\eea}{\end{eqnarray}}

\begin{document}

\preprint{CERN-PH-TH-2010-320}
\preprint{RIKEN-MP-10}

\title{Rapid Thermalization 
by Baryon Injection in 
Gauge/Gravity Duality}

\date{\today}

\author{Koji {\sc Hashimoto}}\email[]{koji(at)riken.jp}
\affiliation{
{\it Mathematical Physics Lab., RIKEN Nishina Center, Saitama 351-0198,
Japan }}

\author{Norihiro {\sc Iizuka}}\email[]{norihiro.iizuka(at)cern.ch}
\affiliation{
{\it Theory Division, CERN, CH-1211 Geneva 23, Switzerland}}

\author{Takashi {\sc Oka}}\email[]{oka(at)cms.phys.s.u-tokyo.ac.jp}
\affiliation{
{\it Department of Physics, Faculty of Science, University of Tokyo, 
Tokyo 113-0033, Japan}}

\begin{abstract}
Using the AdS/CFT correspondence for strongly coupled gauge theories, 
we calculate thermalization of mesons caused by a time-dependent 
change of a baryon number chemical potential. On the gravity side, 
the thermalization corresponds to a horizon formation 
on the probe flavor brane in the AdS throat. Since 
heavy ion collisions are locally approximated by a sudden change of 
the baryon number chemical potential, we discuss implication of our results to 
RHIC and LHC experiments, to find a rough estimate of rather rapid 
thermalization time-scale $t_{\rm th} < 1 \,$[fm/c]. 
We also discuss universality of our analysis against varying gauge theories.
\end{abstract}

\maketitle


\noindent

{\centerline {\bf Introduction}} 
The AdS/CFT correspondence \cite{Maldacena:1997re,Gubser:1998bc,Witten:1998qj}, 
or more broadly, the gauge/gravity duality, is an extremely useful tool 
to study strongly coupled field theories. Recently, this correspondence 
has been applied to various field theory settings, and these applications 
open up many new correspondences between gravity to  
other branches of physics. Perhaps one of the most surprising things in 
the success of using gravity to study strongly coupled gauge theories is that 
it seems to work even for an explanation of 
heavy-ion collision 
experiment data at Brookhaven's Relativistic Heavy Ion Collider (RHIC). 
In RHIC experiments \cite{Shuryak:2003xe,Shuryak:2004cy}, one of the big surprises 
was  that a quark-gluon plasma (QGP)  forms at a very early stage 
\cite{Heinz:2004pj} just after the heavy ion collision, {\it i.e.}  a rapid thermalization occurs. 
This obviously requires a theoretical explanation, 
but remains as a challenge, because this requires a calculation of the strongly coupled field theory in 
non-equilibrium process. In this letter, we study the thermalization in strongly coupled field theories
by using the gauge/gravity duality. 

The key idea is to approximate the heavy ion collision by a 
sudden change of a baryon-number chemical potential locally at the collision point. 
Using the AdS/CFT correspondence, 
we obtain strongly coupled gauge theory calculations 
for the thermalization,  
where a time-dependent confinement/deconfinement transition 
occurs due to a sudden change of the baryon-number chemical potential, 
with dynamical degrees of freedom changing from mesons to quark/gluon thermal plasma.  
We calculate a time-scale for that. 
Our strategy can be summarized briefly as follows;  
On the gravity side of the AdS/CFT correspondence, 
the change in the
baryon chemical potential is encoded in how we throw in the baryonically-charged 
fundamental strings (F-strings) from the boundary to the bulk. 
Since the F-string endpoint is a source term for the gauge fields 
on the flavor brane in the AdS bulk, 
this provides a time-dependent gauge field configuration. 
This induces a time-dependent effective metric for the degrees of 
freedom on the flavor brane, 
which are mesons. As a result, this yields the emergence of an apparent horizon on the flavor brane, which signals, in the dual strongly coupled field theory,  
the ``thermalization of mesons'', which we mean that 
the meson degrees of freedom change into quark and gluon degrees of freedom 
with thermal equilibrium. 

Any computation of thermalization of mesons due to the injection of the 
baryon charge in strongly coupled gauge theories has never been proposed. 
We provide a generic framework for it in this paper.
We also present computations at different gauge theories, and argue how universal  
the thermalization time-scale is. We also discuss 
its implication in strongly coupled gauge theories, which hopefully offers a
path to realistic QCD.  
The observation of the flavor thermalization due to changes of external parameters ({\it i.e.}, quantum quench)
different from the baryon charge, was studied in \cite{Das:2010yw}. 
Previous studies on holographic
thermalization, for example in \cite{Janik:2006gp,Chesler:2008hg,Bhattacharyya:2009uu,Chesler:2009cy,Chesler:2010bi},  
discussed glueball sectors, while ours observes the meson sector thermalization. 
What we see is the thermalization on the probe flavor brane.
Since the back-reaction is not taken 
in our setting, this thermalization is not at all related to the one of the glueball sectors.

In our framework, the only input is the function which represents how 
we throw in the baryonically charged F-strings. Therefore, we have 
small number of parameters, which includes a typical maximum value of the baryon density  and 
the time-scale for changing the chemical potential.  
With collision parameters at RHIC, we obtain the thermalization time-scale as 
$t_{\rm th} < 1 \;[{\rm fm/c}]$. 
Actually this time-scale can be well compared with the known hydrodynamic simulation requirement 
$t_{\rm th} < 2 \; [{\rm fm/c}]$ discussed for example in 
\cite{Hirano:2001eu,Kolb:2000fha,Teaney:2001av,Huovinen:2001xx,Heinz:2002un}.
We also ``predict'' that heavy-ion collisions at CERN's Large Hadron Collider (LHC)  
exhibit slightly smaller order of the time-scale for thermalization as 
$t_{\rm th} \lesssim  {\cal{O}} \left(0.1 \right) \;[{\rm fm/c}]$. 

Let us make a few more comments about comparing our analysis 
with data. As we mention previously, we are discussing the time-scale of 
``thermalization of mesons'', which is the horizon formation on the probe brane only. 
In the real world experiments such as RHIC or LHC, 
thermalization should involve not only mesons but also glueballs, 
which corresponds to the black hole horizon formation in the bulk geometry, 
not only on the probe brane. 
Since we are not treating the glueball sectors, the ``prediction'' above in order to compare our analysis with the data is unfortunately not as accurate as it should be. 
Another worry for comparison with data is that we are discussing the large $N$ limit of gauge theories. 
Therefore, the reader should regard our analysis as just an indication of rapid thermalization of some sectors 
of the large $N$ gauge theories. 

In the following, after solving the equations in the gravity side with
a generic time-dependent baryon 
chemical potential, we compute the apparent
horizon and the time-scale for the thermalization. The simplest 
example offered is 
${\cal N} = 4$ super Yang-Mills with ${\cal N}= 2 $ hypermultiplets as ``quarks''. 
We conclude with
the statement of the universality, by showing some variations of the
setup, including 
quark masses and confining scales.

\vspace{5mm}

{\centerline {\bf D3-D7 system with quark injection}}

\vspace{2mm}

The simplest set-up in AdS/CFT with quarks is the 
${\cal N}=2$ supersymmetric massless QCD constructed by a 
D3-D7 system \cite{Karch:2002sh}, where we consider the gravity back-reaction of only D3-branes and regard 
D7-brane as a prove flavor brane. We
are interested in the dynamics of mesons and the deconfinement of
quarks, which is totally encoded in the probe flavor D7-brane in the
$AdS_5\times S^5$ geometry,
\begin{eqnarray}
 ds^2 = \frac{r^2}{R^2}\eta_{\mu\nu} dx^\mu dx^\nu
+ \frac{R^2}{r^2}
(d\rho^2\! +\! \rho^2 d\Omega_3^2\! +\! dw_5^2\! +\! dw_6^2), 
{\label{adsmetric}}
\end{eqnarray}
where $\rho^2 \equiv w_1^2 + w_2^2 + w_3^2+w_4^2$ and 
$r^2 \equiv \rho^2 + w_5^2 + w_6^2$. $R$ is the AdS radius defined by
$R^4 = 4\pi g_s N_c \alpha'^2$. The string coupling is related to the
QCD coupling as $2\pi g_s\equiv g_{\rm QCD}^2$.
The dynamics of the flavor D7-brane is determined by the D7-brane action
\begin{eqnarray}
 S = -\mu_7 \int d^8\xi
\sqrt{-\det
\left(
G_{ab}
+ 2\pi\alpha' F_{ab}
\right)}\, ,
\label{D7action}
\end{eqnarray}
where the D7-brane is at $w_5 = 0$ and are extended on the gauge theory 
directions $x^{\mu}$ ($\mu = 0, 1, 2, 3$), 
$\rho$, and $\Omega^3$. 
The fluctuations of gauge fields $A_a$ (or, scalar field 
$\eta \equiv w_6$) 
on the D7-brane corresponds to vector (or, scalar) mesons. 
For a concise review of the D3-D7 system and meson dynamics, see
\cite{Erdmenger:2007cm}.
$G_{ab}$ is the induced metric on the
D7-brane. The asymptotic 
$(\rho \to\infty)$ value of $\eta$ corresponds to the quark mass 
$m_q = \eta/2\pi\alpha'$. For simplicity we first put it zero. This corresponding to a 
``marginal confinement'' for the mesons on the flavor brane since it has only zero-sized 
horizon \footnote{Therefore this D7-brane is already touching the zero size (and zero temperature) bulk horizon, which is so-called ``black hole embedding''. Note that this induces zero temperature black hole on the probe D7-brane, which indicates ``marginally confining phase'' because of zero temperature.  
However in this paper, we analyze how the {\it non-zero size} (and non-zero temperature) horizon is formed on this probe brane due to the time-dependent chemical potential change.}.
The D7-brane tension is $\mu_7\equiv 1/ (2\pi)^7 g_s \alpha'^4 $. 

In AdS/CFT, 
the response to the change in the baryon chemical potential is totally
encoded in this D7-brane action. We will solve this gauge field for
arbitrary time-dependent chemical potential. 

As the chemical potential of our interest is
homogeneous, we may turn on only the $F_{tr}$ component. With a
redefinition of the AdS radial coordinate $z \equiv R^2/\rho$, the 
D7-brane action (\ref{D7action}) is equivalent to a 1+1-dimensional
Born-Infeld system in a curved background,
\begin{eqnarray}
 S = -\mu_7 V_3 {\rm Vol}(S^3)
\int\! dt dz
\frac{R^8}{z^5}
\sqrt{
1-\frac{z^4}{R^4}(2\pi\alpha')^2 F_{tz}^2
} \, ,
\label{effectiveaction}
\end{eqnarray}
where $V_3$ is the volume of $x^{1,2,3}$ space. 

\begin{figure}
\centering
\includegraphics[scale=0.4]{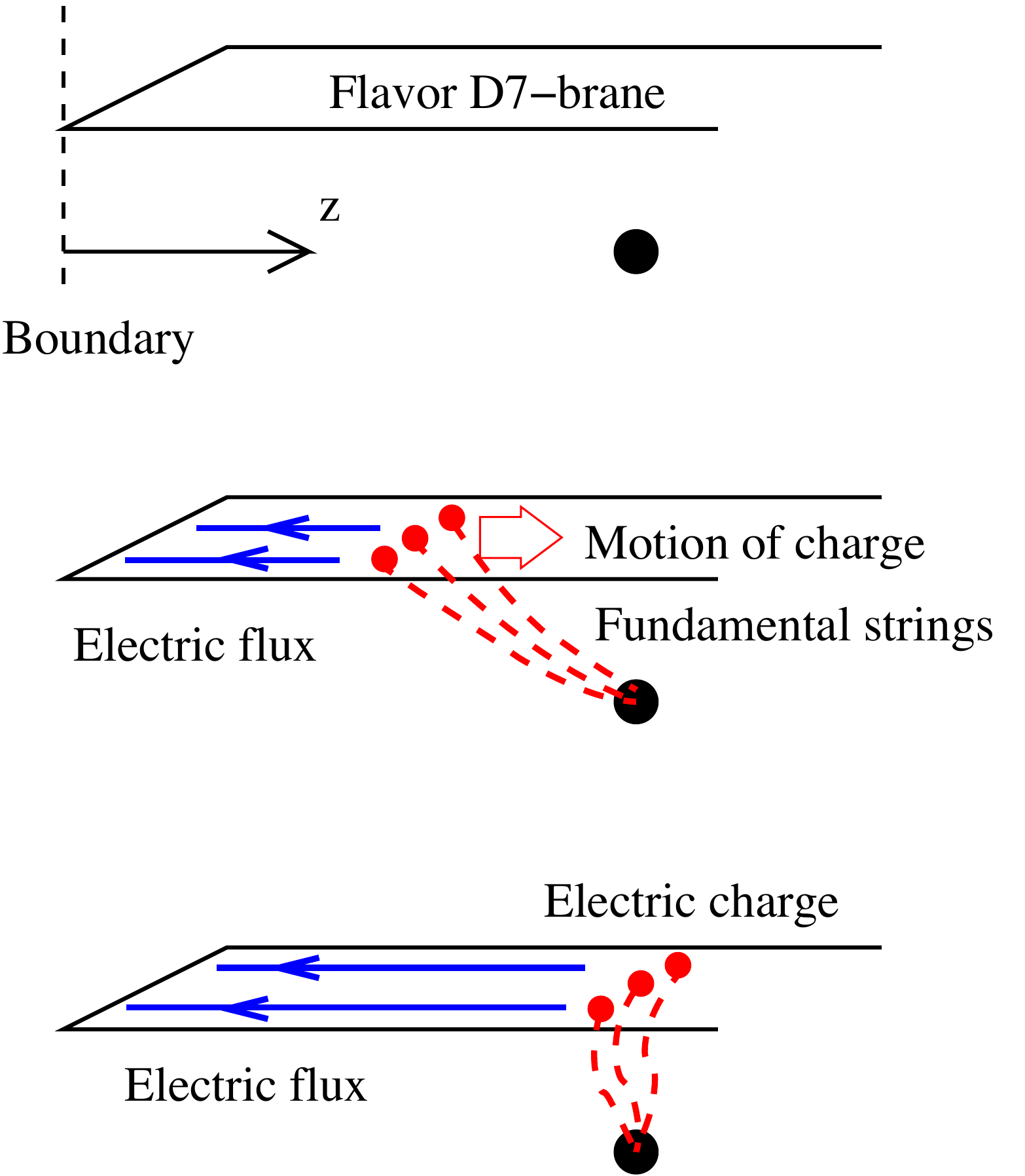}
\caption{How the baryon charge density changes in time, as we throw-in
fundamental strings from the boundary. Top figure: there is no baryon
 charge. Bottom figure: the static 
baryon (quark) charge is provided by the
fundamental strings (dashed lines) 
connecting the probe D7-brane and the AdS horizon (black blob). 
The end points of the fundamental strings are electric
 charges (flavor charges) on the D7-brane. An electric flux 
 (solid lines with arrows) emanates
from those charges on the D7-brane. Middle figure: the flavor electric
 charges are thrown-in from the AdS boundary into the bulk. This
 specifies the time-dependent chemical potential.  The electric flux is time-dependent. 
 The big arrow denotes the motion of the electric charges on the D7-brane.}
\label{figcharge}
\end{figure}

In the AdS/CFT dictionary, the static baryon chemical potential corresponds to 
$A_t(r=\infty)-A_t(r=0)$. This basically counts the number of electric
charges located at the origin $r=0$. In order to change this number in
a time-dependent manner, we need to consider an additional source term
\begin{eqnarray}
 \delta S = \mu_7 V_3 {\rm Vol}(S^3)
\int\! dt dz
\left(A_t j^t + A_z j^z\right) \, ,
\end{eqnarray}
which describes end-points of a fundamental string (electric charges)
thrown in from the outside of the system, {\it i.e.}, from the boundary
into the bulk. See Fig.~\ref{figcharge}.
Since the geodesic of the fundamental string end-points is the light 
geodesics determined by the induced metric $G_{ab}$, and in the
present case $G_{tt} = - G_{zz}$, it is just along a null vector $(v_t, v_z) = (1,-1)$. 
Therefore the source current is
an arbitrary function of the variable $t-z$. With a current conservation
relation, we obtain that 
$j^t = j^z$, and we take the arbitrary source function as  
$j^t = j^z = g'(t-z)$.
Given this source current $j$,  the gauge field strength $F_{tz}$ is readily solved from the equations of motion on the 
D7-brane and we obtain 
\begin{eqnarray}
(2\pi\alpha') F_{tz}= \frac{R^2 z \; g(t-z)}
{\sqrt{(2\pi\alpha')^2 R^{12}+ z^6 \left(g(t-z)\right)^2}} \,.
\label{gaugesol}
\end{eqnarray}
This is the gauge field solution which encodes the information of the
time-dependent chemical potential given the source current $j^t = j^z = g'(t-z)$.

The relation between $g$ and the chemical potential is linear.
For the static case $g(t-z) =$ constant, the solution (\ref{gaugesol})
is nothing but a conventional Born-Infeld solution on
a D-brane. So we can compare the Born-Infeld charge $g$ with the
fundamental string charge (which equals the quark number), following the
techniques found in \cite{Callan:1997kz}, to obtain
\begin{eqnarray}
 g(t) = (2/\pi)(2\pi\alpha')^4 \lambda \, n_{\rm B}(t) \,,
\label{baryon}
\end{eqnarray}
where the baryon number density is $n_{\rm B}=n_{\rm quark}/N_c$ at the 
boundary $z=0$. 
This determines the normalization of the baryon number, in the solution
(\ref{gaugesol}) of the D7-brane system.


\vspace{5mm}
{\centerline {\bf Horizon formation on the flavor D7-brane}}

\vspace{2mm}

A time-dependent configuration on the D7-brane modifies the effective
metric which the fluctuations on the D7-brane feels. A large field
configuration creates a horizon of the metric 
on the D7-brane, which signals the
thermalization in AdS/CFT. From the induced metric with the background 
gauge configuration (\ref{gaugesol}), here 
we compute the location of an apparent horizon on the
D7-brane \footnote{It is a difficult issue to define a temperature in
time-dependent system. Since event horizon is always 
outside of the apparent horizon, as far as apparent horizons do not disappear, 
we regard the 
emergence of an apparent horizon as a signal of thermalization. 
Event horizons cannot be defined locally, while apparent horizons can.  
In static space-time, these two horizons coincide.}.

Let us compute an effective metric which a scalar fluctuation 
$\eta$ feels, which is massless scalar meson like pion field. 
By expanding the D7-brane action (\ref{D7action}) for small 
fluctuation $\delta \eta$ in the background solution $F_{tz}$ (\ref{gaugesol}), to the
quadratic order, we can obtain effective metric $\tilde g$ as:
\begin{eqnarray}
 S 
= - \!\!\int \! dt dz d^3x^{i} d^3\theta^I 
\; \frac{\sqrt{- \tilde{g}}}{2} \tilde{g}^{MN}\partial_M \delta \eta \partial_N \delta \eta + {{\cal O}(\delta \eta^3)},\,
\end{eqnarray}
where $i=1,2,3$ is the spatial directions of our 3+1-dimensional space-time, 
and $\theta^I$ ($I=1,2,3$) is the angular variable on
the $S^3$. $(M,N)$ shows the whole 7+1 directions on D7-branes, $(t,z,i,I)$.
The effective metric $\tilde{g}$ can be obtained easily,
\begin{eqnarray}
 - \tilde{g}_{tt} & =& \tilde{g}_{zz}\nonumber \\
& =&  \mu_7^{1/3} R^{4/3}z^{-4/3} (1-z^4 R^{-4}
(2\pi\alpha'^2)F_{tz}^2)^{5/6}, \\ 
 \tilde{g}_{ij}& =&  \mu_7^{1/3} R^{4/3}z^{-4/3}
 (1-z^4 R^{-4}
(2\pi\alpha'^2)F_{tz}^2)^{-1/6} \delta_{ij}, 
\nonumber
\\ 
 \tilde{g}_{IJ}& =& \mu_7^{1/3} R^{4/3} z^{2/3} 
(1-z^4 R^{-4}
(2\pi\alpha'^2)F_{tz}^2)^{-1/6} G_{IJ},
\nonumber
\end{eqnarray}
where $G_{IJ}$ is the metric on the unit 3-sphere.

Given this effective metric, we will now determine the apparent horizon, which is 
defined locally as a surface whose area variation vanishes along the null rays which is normal to the surface.
The surface area at an arbitrary point in given $(t,z)$ is 
\begin{eqnarray}
 V_{\rm surface} \!\!\!& =&\!\!\! \int d^3x^i d^3\theta^I
\sqrt{
(\Pi_{i=1,2,3} \;\tilde{g}_{ii})
(\Pi_{I=1,2,3} \; \tilde{g}_{II})
}
\\
& =& \!\!
V_3 {\rm Vol}(S^3) \mu_7 R^4 z^{-1}
(1-z^4 R^{-4} (2\pi\alpha'^2)F_{tz}^2)^{-1/2}.
\nonumber
\end{eqnarray}

The $(t,z)$ spacetime has a trivial null vector normal to the surface $(v^t, v^z)=(1,-1)$
since $ \tilde{g}_{zz} = - \tilde{g}_{tt}$,
so the constancy of the surface area variation along this null ray is
\begin{eqnarray}
d V_{\rm surface} |_{d t = - d z} = 0 \,,
\end{eqnarray}
which yields
\begin{eqnarray}
\left( \partial_z -
\partial_t
\right)
\left[z^2 (1-z^4 R^{-4} (2\pi\alpha'^2)F_{tz}^2)\right]=0.
\label{etaapp}
\end{eqnarray}
Substituting the gauge field solution (\ref{gaugesol}) to this, we
obtain the following equation 
\begin{eqnarray}
 (2\pi\alpha')^2 R^{12} - 2 z^6 g^2 + 2 z^7 g g' = 0 \,.
\label{apparent}
\end{eqnarray}
If this equation (\ref{apparent}) admits a solution, it specifies where the apparent horizon on the D7-brane is formed. 

Before we proceed, a few comments are in order. First, 
we are calculating the apparent horizon, not the event horizon on the flavor brane. 
Since apparent horizon is always inside the event horizon, as long as apparent horizon never disappears at finite time, we can regard the formation of the apparent horizon as a signal of the thermalization of system. 
However since the positions of the apparent horizon are time-dependent, it is difficult to 
extrapolate the thermodynamical information such as temperature, which is determined by the event horizon. 
On the other hand, the apparent horizons are defined locally without knowing the late time asymptotics, 
therefore it has a calculation simplicity. 
Secondary, we are using the Born-Infeld action on the flavor D7-brane to determine the effective metric for the 
various mesonic modes. This Born-Infeld form is crucial, since if we have used 
the Yang-Mills form for the D7's degrees of freedom, the horizon would have not 
formed. The reason why we need the Born-Infeld form is due to the warping in the (\ref{adsmetric}), which 
makes the effective string tension $\alpha'$ finite, so it is not appropriate to replace 
the Born-Infeld action by the simple Yang-Mills form. 


\vspace{5mm}
{\centerline {\bf Thermalization time-scale order estimation}}

\vspace{2mm}

From ({\ref{apparent}}), 
we can order-estimate the thermalization time-scale by a dimensional analysis without specifying the 
explicit form of the source function $g(t-z)$.  
We would like to consider the chemical potential change which mimics the heavy-ion collisions. 
The function $g(t-z)$ at the $AdS_5$ boundary $z=0$ is
directly related to the time-dependent baryon
number according to (\ref{baryon}) as 
$g(t) = (2/\pi)(2\pi\alpha')^4 \lambda \,n_{\rm B}(t)$.

Suppose that $g(t-z)$ changes like trigonometric functions from 
zero to some maximal value during the time-scale $1/w$. 
This is like the situation where the chemical potential change locally by
two baryonically-charged heavy ions approaching each other. 
Setting the maximal value of 
$g(t-z)$ as  $g_{\rm max}$,  
we can order-estimate it as
\begin{eqnarray}
g(\xi) \sim g_{\rm max}  \,, \, g'(\xi) \sim w g_{\rm max}  \,.
\end{eqnarray}
Then, a dimensional analysis of (\ref{apparent}) estimates  
the thermalization time-scale $t_{\rm th}$ as 
\begin{eqnarray}
t_{\rm th} \sim \left(\frac{(2 \pi \alpha')^2 R^{12}}{g^2_{\rm max}} \right)^{1/6} 
\sim \left( \frac{\lambda}{n_{\rm B}^2} \right)^{1/6} \,,
\end{eqnarray}
if $t_{\rm th} w \lesssim 1$ is satisfied. Here $n_{\rm B}$  being the maximal baryon number density determined as
$g_{max} = 4(2\pi\alpha')^4 \lambda n_{\rm B}$. 
If $t_{\rm th} w \gg 1$ instead, then we obtain, 
\begin{eqnarray}
t_{\rm th} \sim \left(\frac{(2 \pi \alpha')^2 R^{12}}{g^2_{\rm max} w} \right)^{1/7} 
\sim \left( \frac{\lambda}{n_{\rm B}^2 w} \right)^{1/7} \,.
\end{eqnarray}

On the other hand, if $g(\xi)$ has an explicit $\xi$ dependence like 
a power-law behavior to approach its maximum, such as $g(\xi) \propto \xi^n$, with positive $n$, then 
\begin{eqnarray} 
g(\xi) \sim g_{\rm max} (w \xi)^n  \,, \, g'(\xi) \sim w g_{\rm max} (w \xi)^{n-1} \,. 
\end{eqnarray}
Again, a dimensional analysis yields,   
\begin{eqnarray}
t_{\rm th} \sim \left(\frac{(2 \pi \alpha')^2 R^{12}}{g^2_{\rm max} w^{2n} } \right)^{1/{(6 + 2 n)}} 
\sim \left( \frac{\lambda}{n_{\rm B}^2 w^{2n} } \right)^{1/{(6 + 2 n)}} \,.\quad
\end{eqnarray} 

In summary, in terms of following parameters, inverse of the 
variation timescale of the baryon chemical potential $w$, 
't Hooft coupling $\lambda$, and the maximum baryon density $n_{\rm B}$, 
the thermalization time scale is written as  
\begin{eqnarray}
t_{\rm th} 
\sim \left( \frac{\lambda}{n_{\rm B}^2 w^{k} } \right)^{1/{(6 + k)}} \,.\quad
\label{genericeq}
\end{eqnarray} 
for given $k (\ge 0)$, which is determined by how we change the baryon number chemical potential. 
This is one of our main results.

In the following, we present two explicit examples of the source function $g(\xi)$ and show that
both examples exhibit the generic behavior (\ref{genericeq}). The first example is for 
a baryonic matter formation, and the second is for baryons colliding and passing through each other.


\vspace{5mm}
{\centerline {\bf Example I : Baryonic matter formation}}

\vspace{2mm}

To understand the time-scale in more detail, let us investigate a few explicit examples. 
The first example presented here mimics colliding baryons forming a baryonic matter.
We start with zero baryon density $n_{\rm B} = 0$ and then 
increase it linearly in time, for $0<t<1/w$. 
It reaches the maximum at $t=1/w$ and then it is kept constant. 
This is like two baryonically charged heavy-ion approaching each other, followed by
a formation of a QGP gas with large baryon number. 
Setting $\xi \equiv t - z$, we arrange the function $g$ accordingly as 
\begin{eqnarray}
 g(\xi) = 
\left\{
\begin{array}{ll}
0  & \,(\xi<0) \\
g_{\rm max}\; w \, \xi  &\, (0<\xi<1/w) \\
g_{\rm max}  & \,(1/w<\xi)
\end{array}
\right.
\label{timedepone}
\end{eqnarray}

\begin{figure}
\centering
\includegraphics[scale=0.75]{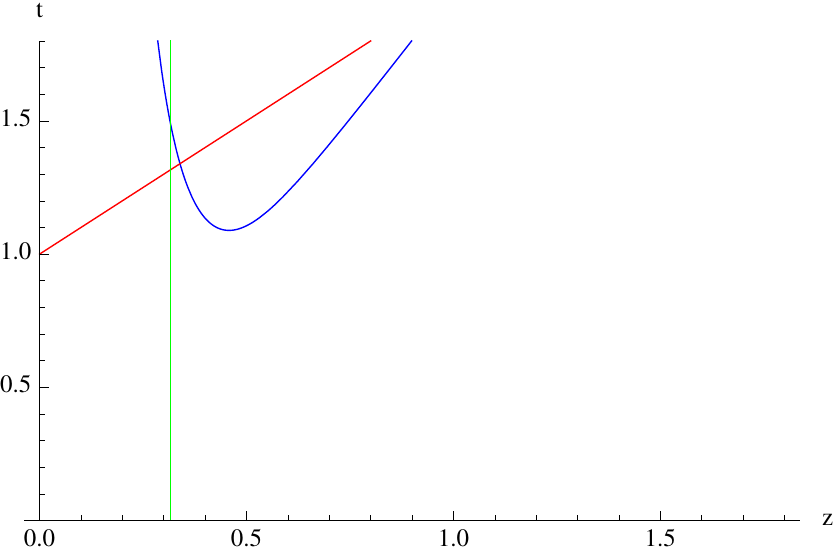}
\caption{Locations of the apparent horizon in $z$-$t$ plane written in the unit of $1/w$ for the parameter is chosen such that 
$(\lambda/(2\pi n_{\rm B}^2))^{1/6} = 1/w$. 
The blue, red, green curves/lines represent the curves/lines (\ref{curveone}), $t = z + 1/w$, and (\ref{curvetwo}), respectively.}
\label{figone}
\end{figure}

With this choice of the time-dependent baryon number density, we can
compute the location of the apparent horizon from (\ref{apparent}). 
The results are,
\begin{eqnarray}
t &=& \frac{3 z}{2} 
+ \frac12 \sqrt{z^2 + \frac{\lambda}{z^6 (2\pi)^4 n_{\rm B}^2 w^2}}
\, \nonumber \\
&&   
\mbox{(for $t < z + 1/w$)}\,,\quad
\label{curveone}
\end{eqnarray}
and
\begin{eqnarray}
z = 
\frac{1}{2\pi^{2/3}}
\left(\frac{{\lambda}}{{n_{\rm B}^2}} \right)^{1/{6}} 
\,\mbox{(for $t > z + 1/w$)} \,.
\label{curvetwo}
\end{eqnarray}
The curve (\ref{curveone}) crosses with line $t = z + 1/w$ at $z = z_0$ where $z_0$ satisfies 
\begin{eqnarray}
1- z_0 w -\frac{ {\lambda}}{64 \pi^4 n_{\rm B}^2 z_0^6} = 0 \,,
\end{eqnarray}
and in the case $z_0 w \ll 1$, $z = z_0$ coincides with the curve (\ref{curvetwo}). 
These results are shown in Fig.~\ref{figone}. 

\begin{figure}
\centering
\includegraphics[scale=0.6]{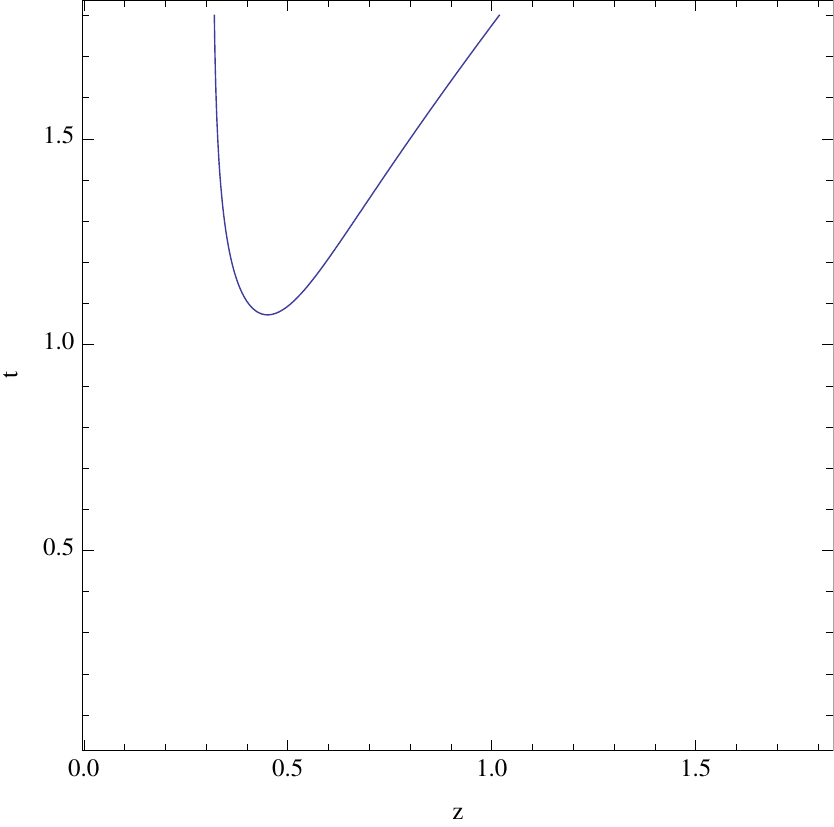}
\caption{A numerical solution for the location of the apparent horizon for the 
source current given by $g(\xi)$ of (\ref{smoothone}), where 
the parameters and the unit are the same as Fig.~\ref{figone}. This is a smoothened 
version of the curve analytically determined in Fig.~\ref{figone}.}
\label{figfive}
\end{figure}
The slight discontinuity between curves (\ref{curveone}) and (\ref{curvetwo}) is
simply due to a cusp of the input source function (\ref{timedepone}) at $\xi = 1/w$. 
If we smoothen the source function (\ref{timedepone}), then the two curves (\ref{curveone}) 
and (\ref{curvetwo}) are connected smoothly. 
Just for a comparison, in Fig.~\ref{figfive} we also show the location curve for the 
apparent horizon for a smooth source 
current given by
\begin{eqnarray}
 g(\xi) &=& 
g_{\rm max}\; \frac{1}{2} \left( 1 + \tanh \left(2  w \, \xi - 1 \right) \right) \nonumber \\
&=& 
\left\{
\begin{array}{ll}
0  & \,(w \xi \ll 1/2) \\
g_{\rm max} \; w \, \xi & \,  (w \xi \sim 1/2) \\
g_{\rm max}  & \,(w \xi \gg 1/2)
\end{array}
\right.
\label{smoothone}
\end{eqnarray}
This Fig.~\ref{figfive} is well compared with Fig.~\ref{figone}.

\begin{figure}
\centering
\includegraphics[scale=0.75]{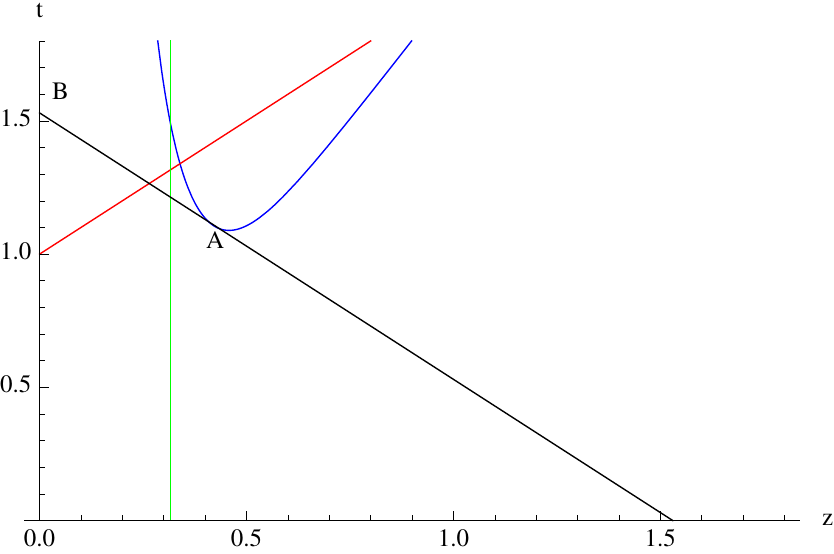}
\caption{
Earliest thermalization for the boundary observer occurs at point A, which is the tangential point between curve (\ref{curveone}) (blue curve) and the light ray propagating toward boundary (black line). The boundary observer see the thermalization at point B. We choose the same paramete as Fig.~\ref{figone}.}
\label{figtwo}
\end{figure}

The emergence of the horizon on the flavor D7-brane is seen by
the boundary observer through a light propagation on the D7-brane.
Suppose that the point A at $(t_{\rm A}, z_{\rm A})$ in Fig.~\ref{figtwo} 
gives the earliest delivery of the information of the apparent horizon. 
Due the fact that the light geodesic toward the $AdS_5$ boundary is again along the null
vector $(v^t, v^z)=(1,-1)$,  
it is clear that the point A is determined by the curve (\ref{curveone}) and its tangential outgoing null line, and it gives  
\begin{eqnarray}
t_A = c_{At} \left({\lambda}/{n_B^2 w^2} \right)^{1/{8}} \,,\, z_{A} = c_{Az} \left( {\lambda}/{n_B^2 w^2} \right)^{1/{8}}  \,, 
\end{eqnarray}
where $c_{At}$ and $c_{Az}$ are order-one coefficients and given by 
$c_{\rm At} \equiv ({11 \sqrt{73}-13})^{1/2} / (48 (2\pi)^4  (5 \sqrt{73}-31 )^3 )^{1/8}
 \approx 0.88$ and $c_{\rm Az} \equiv \left( {(5 \sqrt{73}-31)}/{48 (2\pi) ^4} \right)^{1/8} \approx 0.33$. 
As a result, the thermalization time seen by the boundary observer is 
\begin{eqnarray}
t_{\rm th} &=& t_{\rm A} + z_{\rm A} = (c_{\rm At} \!+\! c_{\rm Az})
\left(\frac{\lambda}{n_B^2 w^2} \right)^{1/{8}} 
\nonumber 
\\
& \sim &
\left(\frac{{\lambda}}{{n_{\rm B}^2 w^2}} \right)^{1/{8}} 
\,.
\label{resultone}
\end{eqnarray}

\begin{figure}
\centering
\includegraphics[scale=0.75]{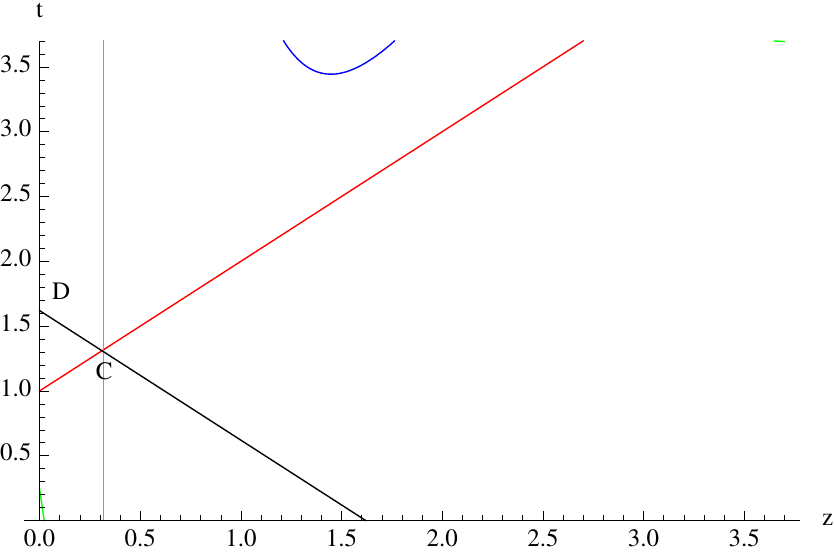}
\caption{Curves for the parameter chosen $(\lambda/(2\pi n_{\rm B}^2))^{1/6} = (10)^{2/3}/w$, 
written in the unit of $1/w$. Thermalization occurs at point C, which is the crossing 
point between line (\ref{curvetwo}) (green line) and the line $t = z + 1/w$ (red line). 
From point C, the light ray is propagating toward boundary (black line), 
and the boundary observer see the thermalization at point D.}
\label{figthree}
\end{figure}

On the other hand, it is also possible that the point C in Fig.~\ref{figthree} gives the earliest occasion, depending
on the values of the parameters in the source function $g(\xi)$. This happens especially if 
\begin{eqnarray}
\frac{1}{w} \ll \left(\frac{\lambda}{n_B^2 w^2} \right)^{1/8} \,.
\label{inequality}
\end{eqnarray}
In this case, similarly 
we can compute the thermalization time $t_{\rm th}$ as 
\begin{eqnarray}
 t_{\rm th} = t_{\rm C}+ z_{\rm C} \,,
\end{eqnarray}
with
\begin{eqnarray}
 t_{\rm C} = \frac{1}{w} + z_{\rm C} \,, 
\quad
z_{\rm C} \equiv  
\frac{1}{2\pi^{2/3}}
\left(\frac{{\lambda}}{{n_B^2}} \right)^{1/{6}} \,,
\label{z0}
\end{eqnarray}
where $z_{\rm C}$ is given by the line (\ref{curvetwo}). 
Therefore, 
\begin{eqnarray}
t_{\rm th} &=& 2 z_C +\frac{1}{w} 
= \frac{1}{\pi^{2/3}}
\left(\frac{{\lambda}}{{n_B^2}} \right)^{1/{6}} + \frac{1}{w} \,. 
\label{resulttwo}
\end{eqnarray}
Due to the inequality (\ref{inequality}), $1/w \ll 2 z_{\rm C}$, this yields 
\begin{eqnarray}
t_{\rm th} &\sim &
\left(\frac{{\lambda}}{{n_B^2}} \right)^{1/{6}} \,. 
\label{resulttwotwo}
\end{eqnarray}

These results (\ref{resultone}) and (\ref{resulttwotwo}) are
consistent with the order estimation (\ref{genericeq}) in the previous section.

\vspace{5mm}
{\centerline {\bf Example II : Baryons passing through each other}}

\vspace{2mm}

\begin{figure}
\centering
\includegraphics[scale=0.75]{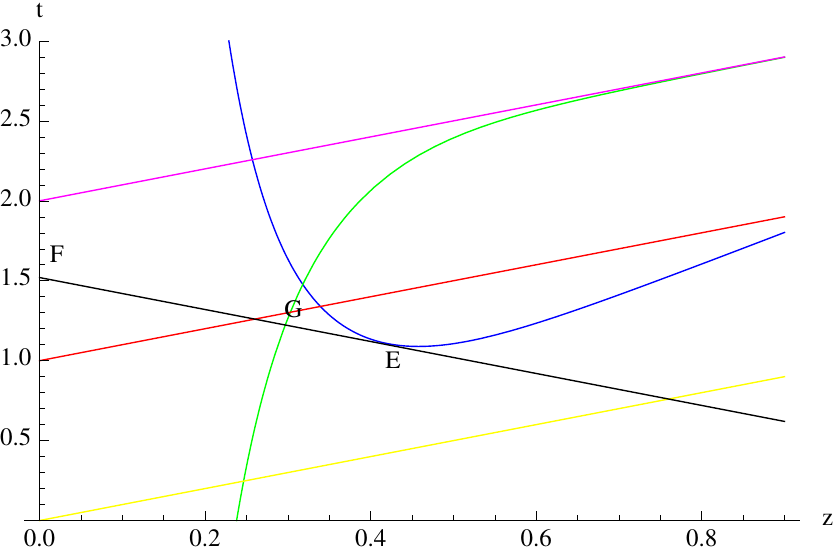}
\caption{The curves (\ref{curveone}) (blue) and (\ref{curvethree}) (green) 
and $t = z$ (yellow), $t = z + 1/w$ (red), $ t = z + 2/w$ (magenta) and 
light ray (black) toward boundary which is tangential to (\ref{curveone}) 
in the unit of $1/w$, where 
the parameter  is chosen such that $(\lambda/(2\pi n_{\rm B}^2))^{1/6} = 1/w$. 
It is clear for this parameter that the earliest apparent horizon is seen at point F, 
which is similar to point B in Fig.~\ref{figtwo}. 
}
\label{figfour}
\end{figure}

Let us investigate another explicit example. This second example mimics
baryons which first collide each other while then pass each other and leave. 
We start with zero baryon density $n_{\rm B} = 0$ and then 
increase it linearly in time, for $0<t<1/w$. 
It reaches the maximum at $t=1/w$ and then decrease to zero again. 
This is like two baryonically charged heavy-ion approaching each other, and then 
they pass by due to the asymptotic freedom. 
Setting $\xi \equiv t - z$, we arrange the function $g$ accordingly as 
\begin{eqnarray}
 g(\xi) = 
\left\{
\begin{array}{ll}
0  & \,(\xi<0) \\
g_{\rm max}\; w \, \xi  &\, (0<\xi<1/w) \\
g_{\rm max} \; (2 - w \xi ) & \,(1/w<\xi<2/w) \\
0 & \,(2/w < \xi) \\
\end{array}
\right.
\label{tdtwo}
\end{eqnarray}
where $g_{\rm max} \equiv (2/\pi)(2\pi\alpha')^4 \lambda n_{\rm B}$ with $n_{\rm
B}$ is the maximum baryon number density. 
Compared with an explicit example I, the location of the apparent horizon for $0<\xi<1/w$ is again given by 
the curve (\ref{curveone}). 
On the other hand, for $1/w<\xi<2/w$, the curve is given by 
\begin{eqnarray}
\label{curvethree}
t &=&\frac{2}{w}+\frac{3 z}{2}  
-\frac{1}{2} \sqrt{z^2 + \frac{\lambda}{z^6 (2\pi)^4 n_{\rm B}^2 w^2}}
 \,\nonumber \\
   && 
   \mbox{(for $z + 1/w < t < z + 2/w $)}\,.\quad
\end{eqnarray}
Finally for $z + 2/w < t $,  (\ref{apparent}) admits no solution, which means that there is no apparent horizon in this region. In Fig.~\ref{figfour}, we plot these curves in the $z$-$t$ plane. 

\begin{figure}
\centering
\includegraphics[scale=0.6]{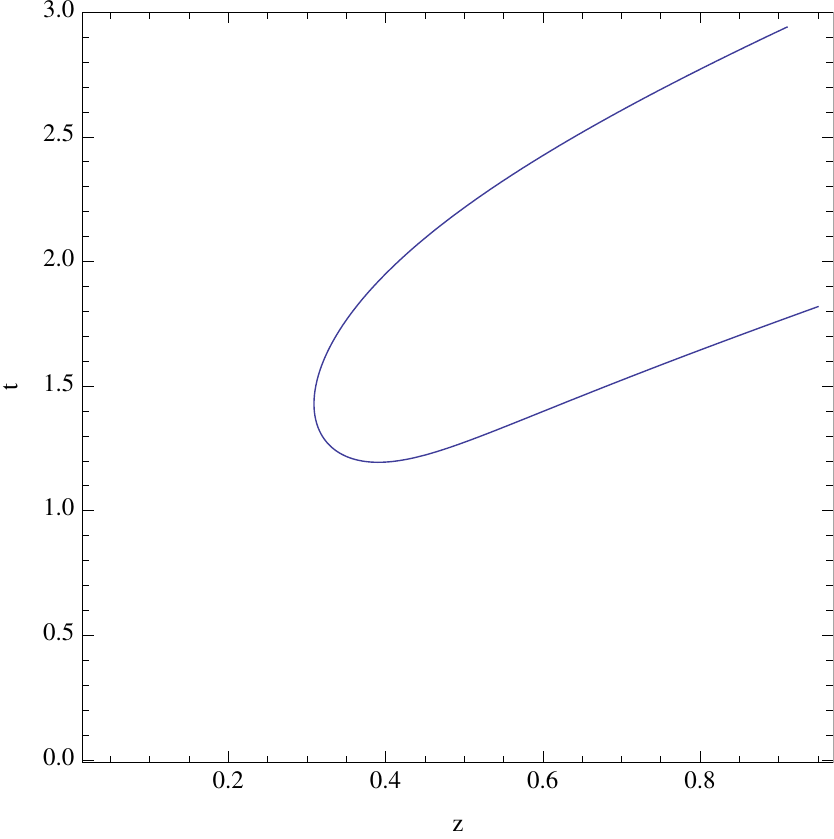}
\caption{A numerical solution for the location of the apparent horizon written 
in the unit of $1/w$ for the source current which is given by $g(\xi)$ of (\ref{smoothtwo}), where 
the parameters are chosen as $(\lambda/(2\pi n_{\rm B}^2))^{1/6} = 1/w$ and written in the unit of $1/w$. 
This is a smoothened version of the curve analytically determined in Fig.~\ref{figfour}.}
\label{figsix}
\end{figure}

Similarly for a comparison, in Fig.~\ref{figsix}  we show also the location curve 
for the apparent horizon for a smooth source 
which takes a Gaussian form as 
\begin{eqnarray}
 g(\xi) &=& 
g_{\rm max}\;  \exp \left( -   \left(2 w\right)^2 \left(\xi - 1/w \right)^2\right) \,.
\label{smoothtwo}
\end{eqnarray}
This Fig.~\ref{figsix} is well compared with Fig.~\ref{figfour}.

Similar to the explicit example I, 
from the curves (\ref{curveone}) and (\ref{curvethree}), we can derive the
thermalization time $t_{\rm th}$. 
From (\ref{curveone}), we obtain again (\ref{resultone}).
From the other curve (\ref{curvethree}), the thermalization point is at the crossing of
the green and the red lines in Fig.~\ref{figfour}. The $z$ value of this crossing point, written as $z_{\rm G}$, is
a solution of the equation
\begin{eqnarray}
\label{zGeq}
\frac{z_{\rm G}^7}{w} + \frac{z_{\rm G}^6}{w^2}  - \frac{\lambda}{4 (2\pi)^4 n_{\rm B}^2 w^2} =0 \, .
\label{eqzg}
\end{eqnarray}
With this, the thermalization time from the curve (\ref{curvethree}) 
is computed as $t_{\rm th} =  2 z_{\rm G}+ 1/w $. 
Therefore, from the two curves, we obtain the thermalization time as the earliest occasion among these two, 
\begin{eqnarray}
t_{\rm th} &=& \mbox{min}
\left\{
(c_{\rm At} + c_{\rm Az}) \left(
\frac{\lambda}{n_{\rm B}^2 w^2} \right)^{1/8} \! \! \! \! \! \!  , \;  
 2z_{\rm G} + \frac{1}{w} \right\} . \quad
\label{minitime}
\end{eqnarray}
Note that in the case where the first term and the third term are well balanced in eq. (\ref{zGeq}), 
it gives, 
\begin{eqnarray}
z_{{\rm G}} \sim \left(\frac{\lambda}{ n_{\rm B}^2 w} \right)^{1/7} \,.
\end{eqnarray}
On the other hand, if the second term and the third term are balanced in eq. (\ref{zGeq}), 
\begin{eqnarray}
z_{{\rm G}} \sim \left(\frac{\lambda}{ n_{\rm B}^2} \right)^{1/6} \,.
\end{eqnarray}
Therefore, if 
\begin{eqnarray}
\label{conditionA}
1/w \lesssim t_{\rm th} 
\end{eqnarray} is satisfied, thermalization time-scale (\ref{minitime}) becomes 
the lowest scale 
\begin{eqnarray}
t_{\rm th} \sim \mbox{min}_{\{k=0,1,2\}} \left\{ 
\left(\frac{\lambda}{ n_{\rm B}^2 w^k} \right)^{1/(6+k)} 
\right\} 
\,.
\end{eqnarray}
Again this is 
consistent with the generic order estimation (\ref{genericeq}). 

\vspace{5mm}
{\centerline {\bf Comparison with experiments}}

\vspace{2mm}

We will later discuss the validity and the 
universality of our results in various other theories which are QCD-like theories.   
But before that, it is quite entertaining to substitute realistic values of the parameters and 
compare our results with the data, even though our setting is not realistic QCD at this point.  
For RHIC and LHC heavy ion collisions, the baryons are passing through each other, so
we may approximate them by the example II above. \footnote{Our calculations just treat homogeneous
change of the baryon chemical potential. However, in heavy ion collisions, various momentum effects 
should take place. So our estimate presented here should be regarded as just an order estimate.}

First, let us consider RHIC parameters.
It is natural to assume that $n_{\rm B}$ is
twice the standard nuclear density $n_{\rm N}$ times the Lorentz contraction factor $\gamma$, 
{\it i.e.}, $n_{\rm B} \sim 2 \gamma n_{\rm N}$, where $n_{\rm N}\sim 0.17 [{\rm fm}]^{-3}$.
In RHIC experiments, we have heavy ion Au-Au collisions with $A = 197$. 
The Lorentz factor is given by the ratio between 
its energy scale $E = \sqrt{s_{NN}}/2 \sim 100 \;[\rm GeV]$ and the mass of Au 
$m_{\rm Au}$, therefore $\gamma = E/m_{Au} \sim 100$. 
On the other hand, at RHIC, the time scale $1/w$ should be given by
the time scale of two nuclei passing by through their bodies, where two 
nuclei are propagating almost with the velocity of light. 
Therefore, $1/w$ is well approximated as $1/w \sim 2 A^{1/3}/ \gamma$ [fm/c], 
where $A$ is the nucleon number and $2 A^{1/3} $ [fm] is the typical nuclear diameter.
This gives $1/w \sim 0.1$ [fm/c]. 

With these inputs at hand, we obtain
\begin{eqnarray}
\left(\frac{\lambda}{w^2 n_{\rm B}^2} \right)^{1/8} 
\sim 
\left(\frac{A^{2/3} \lambda}{\gamma^4 n_{\rm N}^2} \right)^{1/8} 
\sim 0.24 \times \lambda^{1/8} \, [{\rm fm/c}] \,. 
\label{1/8power}
\end{eqnarray}
Similarly, 
\begin{eqnarray}
\left(\frac{\lambda}{w n_{\rm B}^2} \right)^{1/7} 
&\sim& 
\left(\frac{A^{1/3} \lambda}{\gamma^3 n_{\rm N}^2} \right)^{1/7} 
\sim 0.30 \times \lambda^{1/7} \, [{\rm fm/c}] \,, 
\label{1/7power} 
\quad \\
\left(\frac{\lambda}{n_{\rm B}^2} \right)^{1/6} 
&\sim& 
\left( \frac{\lambda}{\gamma^2 n_{\rm N}^2} \right)^{1/6} 
\sim 0.39 \times \lambda^{1/6} \, [{\rm fm/c}]  \,. \quad 
\label{1/6power}
\end{eqnarray}
These are all bigger than $1/w \sim 0.1$, therefore approximately 
(\ref{conditionA}) is satisfied. 
In gauge/gravity duality,  
the 'tHooft coupling  $\lambda = g_{\rm QCD}^2 N_c$ is taken to be very large.
However, {\it since the power of $\lambda$ in $t_{\rm th}$ is small (less than one), 
$t_{\rm th}$ can not take a large value, even for large $\lambda$. 
} 
Actually we use $\lambda\sim {\cal O}(10)$ which is often used in gauge/gravity duality 
for the spectrum comparison. In this case, the smallest of these are given by (\ref{1/8power}), 
though all the scale (\ref{1/8power}), (\ref{1/7power}), and (\ref{1/6power}) give the same order time-scale.   
Therefore we obtain the thermalization time-scale as 
\begin{eqnarray}
 t_{\rm th} < 1 \, [{\rm fm/c}] \,.
\label{thermalizationtime}
\end{eqnarray}

It is interesting that this time-scale can be well compared with the known hydrodynamic simulation 
requirement $t_{\rm th}< 2 \;[{\rm fm/c}]$  
\cite{Hirano:2001eu,Kolb:2000fha,Teaney:2001av,Huovinen:2001xx,Heinz:2002un}.
We found a rather rapid thermalization of mesons.

Our calculation can also gives a ``prediction'' for heavy ion collisions at LHC. 
In LHC where Pb-Pb ion collision experiments are on-going, the energy scale 
is bigger than RHIC as $\sqrt{s_{NN}} = 2.7 \;{\rm TeV}$ \cite{Collaboration:2010bu}. 
The Lorentz factor is $\gamma = E/m_{Pb} \sim 1000$, due to the center of mass difference, and  
therefore, $\gamma$ is 10 times bigger in LHC than RHIC. 
Since  
$A \approx 200$ is almost the same, these give $1/w \sim 0.01 \; [\rm fm/c]$.
With these at hand, we can compute the thermalization time-scale.  Due to the difference of 
Lorentz factor $\gamma$ compared with RHIC case in (\ref{1/8power}), (\ref{1/7power}), (\ref{1/6power}), 
the thermalization time-scale is suppressed furthermore in LHC, and 
we obtain
\begin{eqnarray}
t_{\rm th} \lesssim {\cal{O}} \left(0.1\right)   \;[{\rm fm/c}] \,.
\label{LHCscale}
\end{eqnarray}
Again, with $\lambda\sim {\cal O}(10)$ is used. 
(Due to the asymptotic freedom, $\lambda_{\rm LHC} < \lambda_{\rm RHIC}$, 
however the difference between $ \lambda_{\rm RHIC} $ and $ \lambda_{\rm LHC} $ is tiny therefore 
we can neglect this effect.) 
This (\ref{LHCscale}) gives a significantly faster thermalization time-scale compared 
RHIC.

\vspace{5mm}
{\centerline {\bf Thermalization for various modes}}

\vspace{2mm}

Given the calculation of the scalar meson thermalization in the massless ${\cal{N}}=2$
supersymmetric QCD, it is straightforward to extend the calculation of the thermalization 
for other degrees of freedom. 

First, instead of the thermalization of the scalar meson $\eta$, let us
consider that of vector mesons $A_M$. A similar computation leads to an
effective metric of $A_M$ on the D7-brane, which gives the equation for
the apparent horizon as 
\begin{eqnarray}
\left( \partial_z -
\partial_t
\right)
\left[z^4 (1-z^4 R^{-4} (2\pi\alpha'^2)F_{tz}^2)\right]=0 \,.
\end{eqnarray}
This differs from (\ref{etaapp}) by just a power in the $z$
factor. According to this modification, the thermalization time is just 
$2^{1/3}$ times that of the scalar meson. 
Therefore the thermalization time-scale (\ref{thermalizationtime}) is almost common in order, 
for various vector meson excitations on the flavor branes. 

Next, we consider effects of the quark mass. The quark mass $m_q$
corresponds to the boundary location of the D7-brane, 
$\eta(z\!=\!0) = 2\pi\alpha' m_{\rm q}$. This shifts the D7-brane a bit. 
One can compute the full effect of this shift in our formalism, but
we can give a naive estimate as follows.  
Noticing the fact that $\eta$ comes in the effective metric always as a
combination $(R^2/z^2 + \eta^2/R^2)$ instead of just $R^2/z^2$,
our computation presented here for $\eta=0$ is valid when
\begin{eqnarray}
R^2/z^2 \gg \eta^2/R^2.
\end{eqnarray}
Substituting the expression of $z$ by $z_C$ given by (\ref{z0}), we obtain 
\begin{eqnarray}
 m_{\rm q} \ll (\sqrt{2} \lambda n_{\rm B}/\pi)^{1/3}.
\end{eqnarray}
Even without relying on the large 'tHooft coupling limit, 
this is generally satisfied for light three flavors, for the standard
nuclear density. Therefore again we expect that the 
thermalization time-scale (\ref{thermalizationtime}) is almost common in order, even for the 
various meson excitations with different flavors, such as up, down, and strange flavors.

\vspace{5mm}
{\centerline {\bf Discussion on universality and real-world QCD}}

\vspace{2mm}

We saw that either in massless or massive ${\cal{N}}=2$
supersymmetric QCD, the calculations of the thermalization time-scales of the 
various meson modes are always given by (\ref{genericeq}).  
Given this, it is natural to ask to what extent our thermalization time-scale 
(\ref{genericeq}) holds for a larger variety of gauge theories. 
Since this question is related to a possible universality and also to real-world QCD-related
problems, we shall discuss this question now.

First, note that even though our setting admits supersymmetry, it is also clear that we have never used the 
fermion properties for our thermalization calculations. Therefore we expect that our results are not 
much dependent on the supersymmetry. 

Next, we shall see that, even with different background metrics, 
our thermalization timescale (\ref{genericeq}) is universal.
We have used the  AdS$_5$ metric (\ref{adsmetric}), which represents the deconfined phase for  
the gluon sectors in the conformal ${\cal{N}} = 4$ theory. 
However, the conformality of the gluon sector in the metric (\ref{adsmetric}) is not important at all 
since our computations of thermalization reply only on the asymptotic part 
($z<z_0$) of the induced metric, where $z_0$ is the point where apparent horizon emerge (such as 
$z_{\rm A}, z_{\rm C}, z_{\rm E}, z_{\rm G}$ in our explicit examples I and II). 
Therefore we claim that our results hold for other theories where 
IR dynamics of gluons are significantly different from our theory. 
Even if we replace the metric (\ref{adsmetric}) by some
other non-conformal metric which does not admit a bulk horizon, such as a cut-off $AdS_5$ at 
$z \gtrsim z_{0}$, 
due to the fact that our calculations are insensitive to the IR regime of the geometry at $z \gtrsim z_{0}$, 
our conclusion is still valid. 
In this sense, we expect that our thermalization time-scale (\ref{genericeq}) 
is not only for the ${\cal{N}} = 2$ theory, but rather it works for a broader category of
non-conformal theories.  

The scale $z_0 \lesssim O(1) \;[\rm fm]$ in the example I and II we studied
corresponds to the energy scale $\gtrsim{\cal O} (200) [\rm MeV]$.
Therefore for any non-conformal theory
which admits 
confinement/deconfinement transition for the gluon sectors at the scale smaller than 
$200 [\rm MeV]$, our result (\ref{thermalizationtime}) for the thermalization of mesons is expected to be valid. 
For the real-world QCD, confinement/deconfinement transition is $\sim 200 [\rm MeV]$,  which is the validity bound of our analysis, therefore  
it is expected that our results (\ref{thermalizationtime}) also hold even for realistic QCD. However 
to confirm this, furthermore study is necessary.  

Let us discuss confining gauge theories a bit more. On the gravity side,
confinement is implemented as a deformation at the IR region
($z\sim\infty$) of the geometry. If we use successful hard-wall models in the 
bottom-up models of holographic QCD, 
which is with a cut-off of AdS$_5$ at IR $z > z_0$, then we obtain the same thermalization time-scale.
One other example is a confining geometry made by D3-D(-1) system
\cite{Liu:1999fc,Kehagias:1999iy}, which has the AdS$_5$ form at the UV region. In the solution the
D(-1)'s (D-instantons) condense in the bulk and back-react to modify the  
IR $r \to 0$ region by 
emitting the dilaton $\phi$ as 
\begin{eqnarray}
  ds^2 = e^{\phi/2} \times \left({\rm AdS}_5\times {\rm S}^5 \;
\mbox{geometry}\right), \;
 e^{\phi} = 1 \!+\! \frac{q}{r^4} \,.
\end{eqnarray}
Here $ds^2$ is the metric in the string
frame. In the UV $r \to \infty$, this reduces to the AdS metric (\ref{adsmetric}), as advertized.
The D(-1) charge $q$ (which is proportional to QCD instanton
charge density) is related to the QCD string tension
$\tau_{\rm QCD}$, as
$q/R^8 = \pi^2 \lambda^{-1} \tau_{\rm QCD}^2$. This breaks the
supersymmetries by half.
The universality is valid if this factor $e^\phi$ may not significantly
modify the asymptotic geometry around $z<z_0$, therefore we need to
require $q < R^8 z_0^{-4}$.
With (\ref{z0}), this translates to a condition
\begin{eqnarray}
 \tau_{\rm QCD} < 2^{5/3}\lambda^{1/6}n_{\rm B}^{2/3}.
\end{eqnarray}
Realistic parameters used in this paper show that $\tau_{\rm QCD}$ is
at a comparable order with the right hand side, so this effect may 
modify the thermalization time-scale only slightly by an ${\cal O}(1)$ factor. 
Further study would be interesting for these.  

Finally we comment on possible generalization of our approach for future works. 
We have conducted the calculations with the abelian Born-Infeld action, which treat a single flavor brane. 
In order to treat multi-flavors, we need to 
extend our analysis to non-abelian Born-Infeld action. It is interesting to generalize our study to non-abelian flavor 
branes and study the flavor dependence of the thermalization time-scale.  
Since our thermalization calculations are insensitive to the IR but are sensitive to the UV regime, 
if we consider totally different bulk geometries which do not approach the AdS$_5$ metric (\ref{adsmetric}), 
our result (\ref{genericeq}) is not valid any more. It is quite interesting to generalize our approach to other 
theories where their UV geometries are different from ours, 
such as the holographic QCD model by D4-D8 on Witten's geometry 
\cite{Witten:1998zw,Sakai:2004cn}, or Lifshitz type of 
geometries \cite{Kachru:2008yh} for an application to condensed matter systems.  
We leave these studies for future works.


\vspace{1mm}
\noindent

{\bf --- Acknowledgment.}
K.H.~would like to thank Tetsuo Hatsuda, Yoshitaka Hatta, 
Tetsufumi Hirano, Tadashi Takayanagi, and Koichi Yazaki for helpful comments and  discussions. 
N.I. would like to thank Jorge Casalderrey-Solana, 
Sumit Das, and Kyriakos Papadodimas for helpful discussion and also for  
comments on the draft. 
N.I.~also thanks RIKEN for its hospitality while visiting Japan. 
K.H.~is partly supported by
the Japan Ministry of Education, Culture, Sports, Science and
Technology.

\end{document}